\documentclass[conference]{IEEEtran}
\renewcommand{\baselinestretch}{0.95}
\usepackage{blindtext, graphicx}
\usepackage{cite}
\usepackage{array}
\usepackage{multirow}
\usepackage{amssymb,mathtools}
\usepackage{color}
\usepackage{algorithm}
\usepackage[algo2e]{algorithm2e}
\usepackage{subfigure}
\usepackage{amsmath}

\newcolumntype{P}[1]{>{\centering\arraybackslash}p{#1}}
\def\BibTeX{{\rm B\kern-.05em{\sc i\kern-.025em b}\kern-.08em
    T\kern-.1667em\lower.7ex\hbox{E}\kern-.125emX}}
\begin{document}

\title{Data Driven Optimization of Inter-Frequency Mobility Parameters for Emerging Multi-band Networks}

\author{\IEEEauthorblockN{Muhammad Umar Bin Farooq\IEEEauthorrefmark{1},
Marvin Manalastas\IEEEauthorrefmark{1}, Waseem Raza\IEEEauthorrefmark{1},
Aneeqa Ijaz\IEEEauthorrefmark{1},\\
Syed Muhammad Asad Zaidi\IEEEauthorrefmark{1},
Adnan Abu-Dayya\IEEEauthorrefmark{2} and
Ali Imran\IEEEauthorrefmark{1}}
\IEEEauthorblockA{\IEEEauthorrefmark{1}AI4Networks Research Center, University of Oklahoma-Tulsa, USA}
\IEEEauthorblockA{\IEEEauthorrefmark{2}Department of Electrical Engineering, Qatar University, Doha Qatar\\
\texttt{\{umar.farooq,marvin,waseem,aneeqa.ijaz,asad,ali.imran\}@ou.edu, adnan@qu.edu.qa}}
}

\maketitle

\begin{abstract}
Densification and multi-band operation in 5G and beyond pose an unprecedented challenge for mobility management, particularly for inter-frequency handovers. The challenge is aggravated by the fact that the impact of key inter-frequency mobility parameters, namely A5 time to trigger (TTT), A5 threshold1 and A5 threshold2 on the system’s performance is not fully understood. These parameters are fixed to a gold standard value or adjusted through hit and trial. This paper presents a first study to analyze and optimize A5 parameters for jointly maximizing two key performance indicators (KPIs): Reference signal received power (RSRP) and handover success rate (HOSR). As analytical modeling cannot capture the system-level complexity, a data driven approach is used. By developing XGBoost based model, that outperforms other models in terms of accuracy, we first analyze the concurrent impact of the three parameters on the two KPIs. The results reveal three key insights: 1) there exist optimal parameter values for each KPI; 2) these optimal values do not necessarily belong to the current gold standard; 3) the optimal parameter values for the two KPIs do not overlap.  We then leverage the Sobol variance-based sensitivity analysis to draw some insights which can be used to avoid the parametric conflict while jointly maximizing both KPIs. We formulate the joint RSRP and HOSR optimization problem, show that it is non-convex and solve it using the genetic algorithm (GA). Comparison with the brute force-based results show that the proposed data driven GA-aided solution is 48x faster with negligible loss in optimality. 
\end{abstract}


\IEEEpeerreviewmaketitle

\section{Introduction}
Network densification exploits spatial reuse to increase the network capacity and coverage by deploying a dense heterogeneous network of macro and small base stations (BSs). On the other hand, moving to higher frequency bands also requires reducing cell sizes and concurrent operation at multiple frequency bands \cite{6963801}. However, one caveat of deploying such a huge number of base stations operating on a motley of frequency ranges is the increase in the complexity of the mobility management as well as more pronounced effect of misconfigured mobility parameters on user experience and resource efficiency.  This is due to the proportional increase in the number of handovers (HO), with the increase in the number of BS. It is imperative for the emerging and future networks to have an optimal mobility management as there is a wide range of key performance indicators (KPIs) that directly hinge on user experience and network signaling overhead during handovers. A poor HO management leads to the degradation in several KPIs including data rates, latency, retainability, and user quality of experience (QoE). Optimal HO performance is particularly vital to support Ultra-Reliable Low-Latency Communication (URLLC) use case in 5G \cite{ji2018ultra}.

The current industrial practice of optimizing mobility-related KPIs involves the manual tuning of HO related configuration and optimization parameters (COPs). These COPs are tuned by leveraging domain knowledge and sometimes based on hit and trial approach. In addition to a large number of base stations, an increase in the number of COPs per site emerging network compared to legacy networks makes the problem even more complex. Therefore, with current industry practice manual hit and trial based  COP tuning, managing handovers in the future network is not viable. State of the art Self organizing network (SON) solutions do provide some automation in COP tuning and KPI optimization. For instance, mobility robustness optimization (MRO) is one of the SON functions, which deals with HO parameter management. MRO automatically adjusts a parameter called cell individual offset (CIO) based on the past HO performance of a BS. Though one step ahead of the manual tuning, the current SON solutions would be insufficient for the emerging and future networks due to being reactive and relying on only past observations instead of complete system behavior models \cite{manzoor2020leveraging}. In addition, SON solutions tap on a very limited number of mobility COPs to optimize the KPIs. An efficient and robust HO management can only be devised if the COP-KPI relationship can be quantitatively modeled. However, a tractable analytical COP-KPI model is not feasible to derive due to the system level dynamics and complexity of the cellular network involving mobile users. This calls for investigating data driven models instead.

Data-driven models can be leveraged to quantify the COP-KPI relationship. However, an efficient data driven model needs training data with the following two underlying conditions: 1) data should be sufficiently large and 2) data should be representative. Although, massive data can be mined from a real network meeting the first condition efficiently, the real challenge lies in the representativeness of that data. Aside from the privacy concerns from the subscribers, the main reason for the lack of representative data is the valid reluctance of network operators to test all COP combinations in the live network. To address the issue, in this study we generate and exploit reliable synthetic data to solve the important problem of key mobility parameters optimization for inter-frequency handovers.

\subsection{Related Work}
\label{sec:relatedWork}



A handover is triggered by defined actions called "Measurement Events". 3GPP release 16 \cite{3gpp} has defined standard events for 5G NR which can be used to aid HO decision. Most of the studies optimize HO related parameters of event A3 to improve certain KPIs \cite{tesema2016evaluation,nguyen2017mobility,alhammadi2019auto,nguyen2020geometry,goyal2019handover}.
In \cite{tesema2016evaluation}, authors used CIO as COP to develop a context aware MRO solution for reduction in connection failures using event A3. The authors in \cite{nguyen2017mobility} extended the idea by using three COPs, time to trigger (TTT), offset of event A3 and CIO, to develop a distributed MRO algorithm to minimize radio link failures (RLF). Authors in \cite{alhammadi2019auto} expanded the analysis to 5G settings and used handover margin (HOM) and TTT as COPs while considering user speed and RSRP. They proposed an auto-tuning algorithm to optimize number of handovers and HO failure ratio using event A3. While all the previous studies considered a trade off between ping pong and RLF, the authors in \cite{nguyen2020geometry} extended the state of the art by proving that optimal settings of A3 exist for minimizing both ping pong rate and RLF. In contrast to the previous work, authors in \cite{goyal2019handover} studied the implication of using AHP-TOPSIS method from WiMax for target BS selection in LTE-Advanced cellular networks. They used Q-learning to find optimal value of TTT and hysteresis of event A3. Perhaps, the only study which ventured beyond event A3 was performed in \cite{chaudhuri2017self}. 

In \cite{chaudhuri2017self}, a weighted sum optimization of HO failure ratio, call drop ratio and ping pong ratio using reinforcement learning is done. The study considered TTT and HOM for events A1, A2, A3, A4 and A5. However, this study considers same TTT and HOM for all the events instead of optimizing distinct values of TTT and HOM for different events. 

\begin{table} [!t]
\centering
\caption{{Use of Measurement Events for Inter-Frequency HO}}
\label{table:eventUse}
\begin{tabular}{|>{\centering\arraybackslash} m{2.3cm} | >{\centering\arraybackslash}m{1.5cm}|>{\centering\arraybackslash} m{0.7cm} | >{\centering\arraybackslash} m{0.7cm}| >{\centering\arraybackslash} m{0.7cm}|}
\hline
\textbf{Function} & \textbf{Measurement Event} & \textbf{Vendor 1} & \textbf{Vendor 2} & \textbf{Vendor 3}\\ \hline
\multirow{3}{*}{Inter-Frequency HO} &$A3$&Yes&No&Yes\\\cline{2-5}
 &$A4$&No&Yes&No\\\cline{2-5}
 &$A5$&Yes&Yes&Yes\\\hline
\end{tabular}
\vspace{-0.2in}
\end{table}

The discussed literature investigates intra-frequency handovers using event A3. There is second type of handover called inter-frequency handover, which happens between cells operating on different frequencies. These handovers are more challenging to manage and lead to more signaling overhead and quality of experience issues. Data collected from a leading operator in the United States, operating with 6 frequency bands, show that there are around 60\% more inter-frequency HO attempts compared to intra-frequency HO. This percentage is likely to increase as the number of bands being used increase e.g. due to co-existence of 4G and 5G at different bands. This signifies the importance of inter-frequency HO for current and future cellular networks. However, despite their significance and associated open challenges, inter-frequency HO parameters optimization remains under explored in literature. Table \ref{table:eventUse} shows that all vendors support event A5 for inter-frequency HO, making it the best choice for a self-optimization solution that will work across all the vendors. The superiority of event A5 for managing inter-frequency HO also stems from the increased flexibility and control it offers over the HO execution conditions. With the right settings of A5 parameters such as threshold1 and threshold2, HO can be made sure to happen only at the cell edge. In contrast, A3 that checks only the relative difference between the source and target BS, can lead to handovers even in the middle of cells and is more prone to ping-pongs. However, to the best of the authors' knowledge, there does not exist a study in literature that investigates optimal configuration of A5 parameters for inter-frequency HO.

\subsection{Contributions}
The main contributions of this work are listed below:
\begin{enumerate}
    \item This paper is the first study to quantitatively investigate the impact of key event A5 parameters such as threshold1, threshold2 and TTT on several KPIs such as RSRP and handover success rate (HOSR) for inter-frequency HO. The insights drawn from this analysis show that state of the art gold standard based fixed parameter configuration are not necessarily optimal. A formulation and solution of a multi-KPI maximization problem is required to determine the optimal values of the three parameters.
    \item We formulate and solve a multi-objective optimization problem to determine the optimal values of threshold1, threshold2 and TTT that jointly maximize the RSRP and HOSR. 
    \item To overcome the system level complexity problem that prohibit analytical modeling, we leverage data driven modeling. We evaluate several state of the art machine learning techniques for their potential to generate a reliable COP-KPI model. Results show an XGboost based model outperforms others with less than 2.5\% in HOSR and 0.074dBm in RSRP compared to ground truth. 
    \item To resolve the parametric and objective conflicts observed in the multi-parameters multi-objective optimization problem, we perform Sobol variance based sensitivity analysis. The insights drawn from this analysis are useful for industry for obtaining desired level of gains in HOSR without having to compromise RSRP and vice versa. 
\end{enumerate}

The rest of the paper is organized as follows: Section \ref{sec:systemModel} describes the system model; the qualitative impact of inter-frequency COPs on KPIs is presented in Section \ref{sec:impact}; Section \ref{sec:ML} discusses the performance of machine learning algorithms in capturing the quantitative COP-KPI relationship; Section \ref{sec:optimization} presents the KPI optimization using the machine learning models while Section \ref{sec:conc} concludes the paper.

\section{System Model}
\label{sec:systemModel}
This section describes the 3GPP defined measurement event A5 together with the parameters to optimize the mean RSRP and HOSR. We then formulate the COP-KPI optimization problem and lastly describe the simulation setup for data generation.

\subsection{Handover Event A5}
Event A5 is triggered when RSRP of a user $u$ from serving gNB becomes less than A5 threshold1 and RSRP of the user from target gNB becomes greater than the A5 threshold2. A HO is triggered using event A5 if the following conditions are fulfilled and maintained until TTT is exhausted:
\begin{equation}
\begin{gathered}
    \eta_s^u + A5_{hyst} < A5_{th1}\\
    \eta_t^u + O_{s,t} - A5_{hyst} > A5_{th2}
\end{gathered}
\label{eq:A5_entering}
\end{equation}
where $\eta_s^u$ is the RSRP of the user with serving gNB, $\eta_t^u$ is the RSRP of the user with target gNB, $O_{s,t}$ is the cell specific offset also known as CIO from the serving to target gNB, $A5_{hyst}$, $A5_{th1}$ and $A5_{th2}$ are the hysteresis, threshold1 and threshold2 for event A5, respectively. 

\subsection{Problem Formulation}
RSRP of the user is an important performance metric because it gives an estimate of the link quality between user and the serving BS. Serving cell RSRP also impacts other KPIs such as signal to noise and interference ratio (SINR) and throughput. The downlink RSRP $\eta_s^u$ for a user $u$ connected to the serving BS $s$ is given by:
\begin{equation}
    \eta_s^u = P_s d_s^u
\label{eq:RSRP}
\end{equation}
where $P_s$ is the transmit power of serving BS $s$ and $d_s^u$ is the pathloss dependent component of the user $u$ with the serving BS $s$.
Poor settings of HO parameters can impact the RSRP of the user, i.e., a very high value of TTT can cause too late HO keeping the user in inadequate RSRP for a long time. Similarly, a bad settings of A5 thresholds, i.e., extreme setting of -90dBm and -120dBm for $A5_{th1}$ and $A5_{th2}$, respectively, will result in too early HO where UEs will be forced to move to BS with much lower RSRP. The mean RSRP $\eta$ of all the users in the network can be described as:
\begin{equation}
    \eta =  \frac{\sum\limits_{\forall i\in U} \eta_s^i}{|U|}
\label{eq:RSRP_network}
\end{equation}
where $U$ is a set of all the users in the network.

HOSR is another important KPI that captures the effectiveness of the HO related parameter settings. It is important to set the A5 parameters to minimize HO failures as poor HOSR increases the signaling overhead, prolongs the user in unsatisfactory signal conditions and can lead to radio link failures. This will ultimately worsen the QoE of user. In addition, the poor HOSR can become a key bottleneck for URLLC in 5G and beyond particularly for applications such as intelligent transport systems and autonomous cars. HOSR $\xi$ can be described as:
\begin{equation}
    \xi = \frac{HOS}{HOS+HOF} \times 100\%
\label{eq:HOSR}
\end{equation}
where $HOS$ and $HOF$ are the number of successful and failed handovers, respectively, in the network.

Mean RSRP and HOSR for the network can be maximized jointly. We formulate a multi-objective optimization problem to maximize $\eta$ and $\xi$ using A5 related COPs as follows:

\begin{equation}
\begin{array}{clcl}
\displaystyle \max_{A5_{TTT}, A5_{th1}, A5_{th2}} &  \alpha\eta_{norm}+(1-\alpha)\xi_{norm}; \\
\textrm{subject to} & T_{min} \leq  TTT \leq T_{max}  \\
& T1_{min} \leq  A5_{th1} \leq T1_{max}  \\
& T2_{min} \leq  A5_{th2} \leq T2_{max}  \\
\end{array}
\label{eq:optimization}
\end{equation}
where $\eta_{norm}$, $\xi_{norm}$ are the normalized values of RSRP and HOSR, respectively and $\alpha$ is the factor that can be used to adjust the relative importance of each KPI. $T$, $T1$, $T2$ are the ranges of TTT, $A5_{th1}$ and $A5_{th2}$, respectively with the subscript showing the minimum and maximum values. The optimization variables are TTT of A5, $A5_{th1}$ and $A5_{th2}$. The three constraints in \eqref{eq:optimization} limit the values of the optimization variables i.e. COPS in the 3GPP defined ranges.

Solving this problem using analytical method is not a viable approach as tractable models for RSRP and particularly HOSR as a function of the three COPs are very difficult if not impossible to derive. Even if abstract mathematical models are created, they cannot capture the dynamics caused by mobility of users. Therefore, to enable practical self-optimization solutions, as originally proposed in \cite{6963801}, data driven modeling is a more viable approach to solve \eqref{eq:optimization}.  

\begin{table}[t]
\centering
\caption{Description of Simulation Parameters}
\label{table:simParameters}
\begin{tabular}{|l|l|}
\hline
\textbf{Parameter Description}                       & \textbf{Value}                 \\ \hline
Number of Macro BS                          & 2                     \\ \hline
Number of Small Cells per Macro BS                      & 1                    \\ \hline
Macro BS and small BS height     & 30m and 20m   \\ \hline
Macro BS and small BS transmit power     & 30dBm   \\ \hline
Total bandwidth for 1.7, 2.1 and 3.5 GHz       & 10, 15 and 20 MHz                \\ \hline
Total PRBs for 1.7, 2.1 and 3.5 GHz       & 52, 78, 106                \\ \hline
Pathloss Exponent  & 3  \\ \hline
Shadowing Standard Deviation    &   4   \\ \hline
User density $\lambda_u$                      & 15 per km$^2$                    \\ \hline
Speed Vector $V$                      & [3, 60, 120, 240] km/h                    \\ \hline
\end{tabular}
\vspace{-0.2in}
\end{table}


\subsection{ Data Generation}
Collecting all the needed data from a live network though plausible in theory, is impractical in practice. This happens because operators cannot afford to try all possible combinations of COPs on live network due to the inherent risk of performance loss during the process. Secondly, such data cannot be shared with academia for privacy and business protection reasons. Even if painstakingly gathered and shared, irrespective of the volume, experience shows in case of cellular networks that real data alone is not representative enough to train reliable models and it has to be augmented with authentic synthetic data anyway.

In this backdrop, to generate the data, we exploit a state of the art 3GPP-compliant system level simulator named SyntheticNET \cite{9084113}. This is the first simulator to model 5G mobility parameters in detail needed for this study. As shown in \cite{9084113}, this simulator has been calibrated against real network measurements to ensure the authenticity of the data generated through it.

A network with an area of size 2km$\times$2km is used for the data generation. We consider a three-tier heterogeneous network where 2 layers are composed of macro cells and the remaining layer is composed of small cells. Each macro cell has three sectors and each sector operates at two frequency bands, 1.7GHz and 2.1GHz. Small cells have omni-directional antenna operating at frequency band of 3.5GHz. The initial deployment of the users in the network follows a uniform distribution with user density $\lambda_u$. Each user can move in the network with speed $v_u$ chosen from a set $V$. All elements of the set $V$ are equally probable and the speed value remains constant for a user. User mobility type is random way point. 
The network level simulation parameters are summarized in Table \ref{table:simParameters} and Table \ref{table:COPs} shows the ranges of event A5 related COPs used to generate the data. Such a wide range for $A5_{th1}$ and $A5_{th2}$ are chosen to cover the effect of hysteresis for making event A5 parameters optimization more robust.

\begin{table}[t]
\centering
\caption{Description of COPs to generate the KPIs}
\label{table:COPs}
\begin{tabular}{|>{\centering\arraybackslash} m{2cm} | >{\centering\arraybackslash} m{4cm}|}
\hline
\textbf{COPs}                       & \textbf{Values}                 \\ \hline
$A5_{TTT}$      & [64, 128, 256, 320, 512] ms                     \\ \hline
$A5_{th1}$   & [-90 to -120] dBm \\ \hline
$A5_{th2}$    & [-90 to -120] dBm \\ \hline
\end{tabular}
\vspace{-0.1in}
\end{table}

\begin{figure}[t]
\centerline{\fbox{\includegraphics[width=0.48\textwidth,height=2.1in]{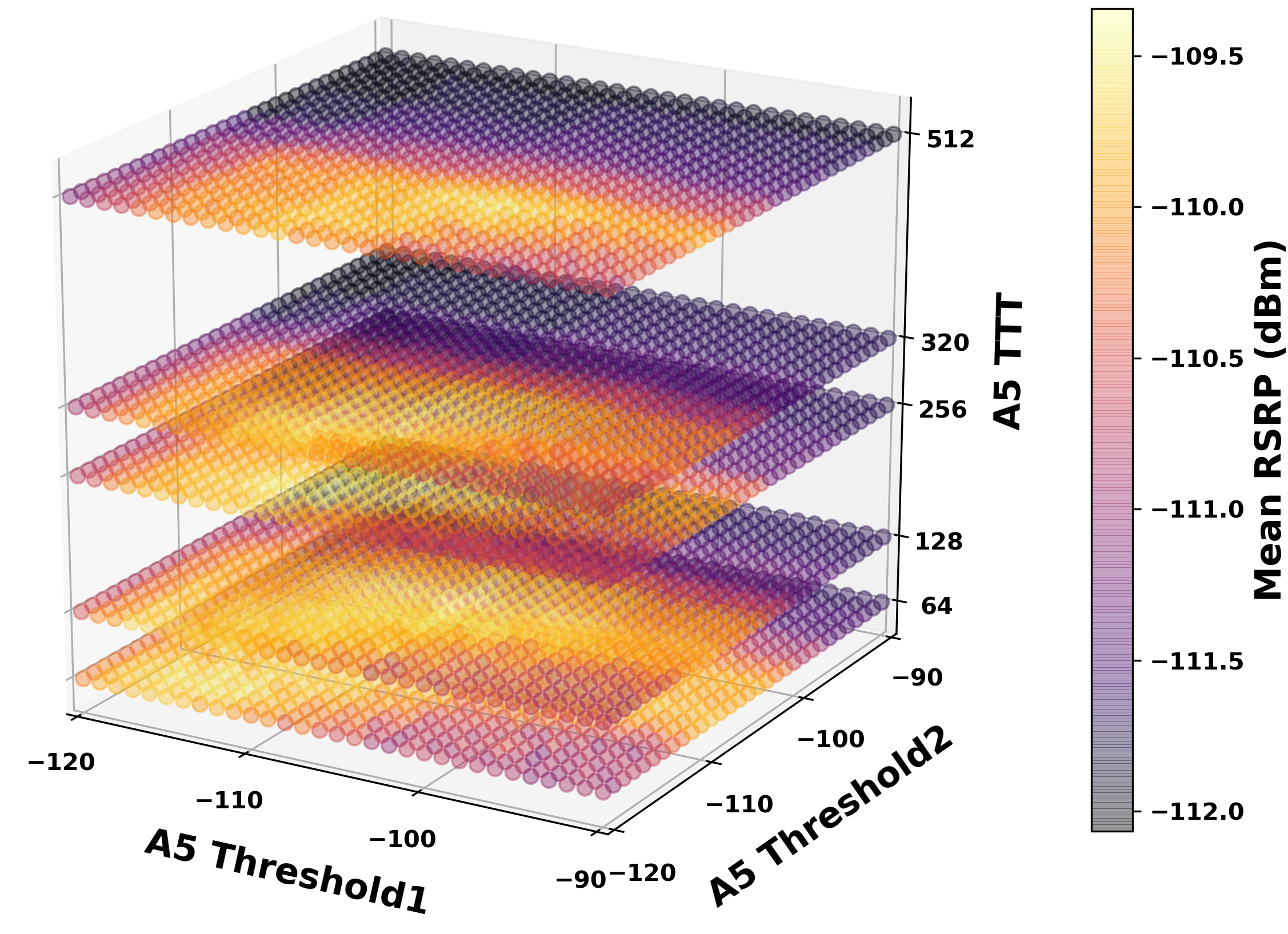}}}
\caption{Impact of A5 thresholds and TTT on mean RSRP.}
\label{fig:RSRP_Thres1vsThres2vsTTT_new}
\vspace{-0.1in}
\end{figure}

\section{Impact of Inter-Frequency Handover Parameters on KPIs}
\label{sec:impact}
To date, the effect of changing the values of A5 related COPs on the KPIs such as RSRP and HOSR is not fully understood, even in academic literature \cite{sim18}. Industry practice on the other hand is to use gold standard fixed values recommended by the vendors for A5 parameter settings without any consideration of their optimality. Qualitative and quantitative insights into how A5 parameter values affect the KPIs are essential to optimize these parameters. This section presents the analysis to harness these insights. These insights are also used to establish the structure of (\ref{eq:optimization}) to see whether or not it is a convex optimization problem so an appropriate solution approach can be adapted.

\begin{figure}[t]
\centerline{\fbox{\includegraphics[width=0.48\textwidth,height=1.5in]{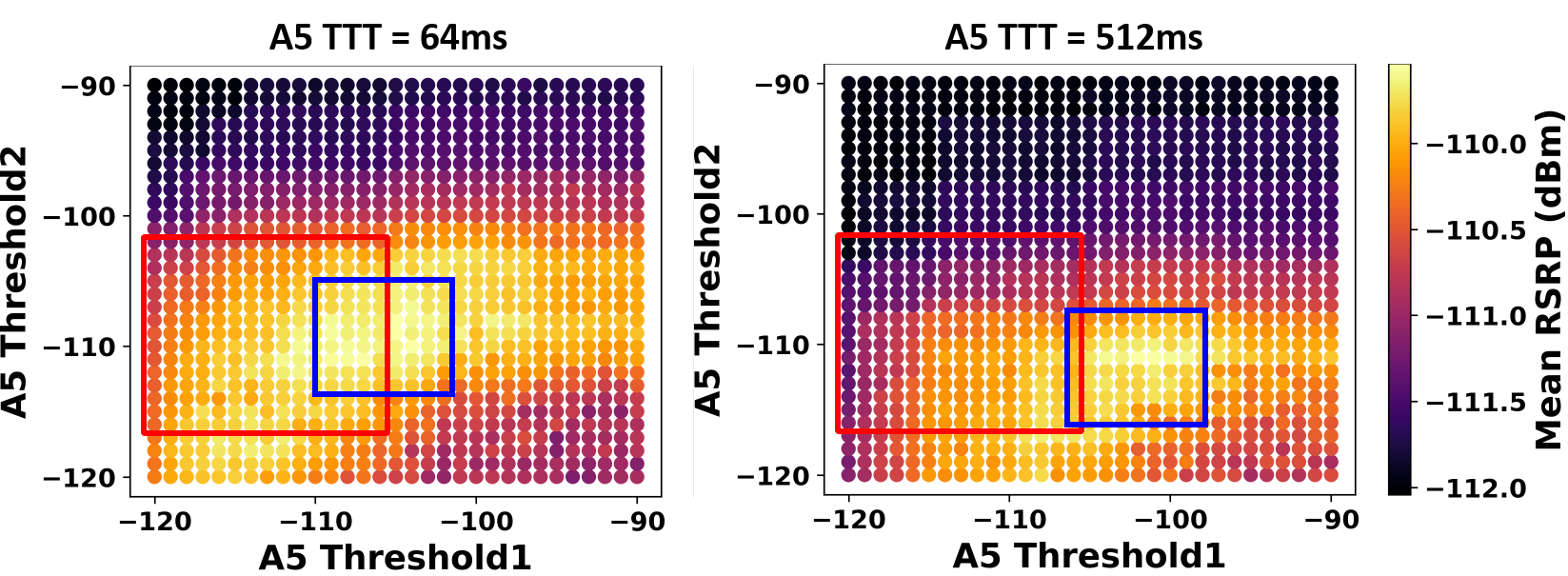}}}
\caption{Comparison of gold standard and simulation results. Red box represents the range of threshold values recommended by the gold standard used in industry. Blue box is the area of high mean RSRP, for analyzed network scenario. This finding can be insightful for the industry.}
\label{fig:GS}
\vspace{-0.1in}
\end{figure}

\subsection{Impact on Mean RSRP}
We begin by analyzing the impact of A5 TTT, $A5_{th1}$, and $A5_{th2}$ on mean RSRP by changing the values of three A5 parameters and logging resultant mean RSRP. Result in Fig. \ref{fig:RSRP_Thres1vsThres2vsTTT_new} shows that the mean RSRP decreases when $A5_{th2}$ values are on the extreme sides. This happens because very high values of $A5_{th2}$ trigger late HO as users are unable to move towards the target BS due to a very high threshold. This ultimately results in a longer stay of users under the coverage of a BS with poorer RSRP. Similarly, lower values of $A5_{th2}$ result in the too early HO to BS with bad coverage lowering the overall RSRP. An opposite effect is observed for variations in the values of $A5_{th1}$. Unlike in $A5_{th2}$, very low values of $A5_{th1}$ cause too late HO as event A5 is triggered when the serving RSRP is already very poor. Meanwhile, very high values of $A5_{th1}$ result in too early HO. In terms of variations in TTT, it is observed that different TTT values shift the high RSRP area. As TTT increases, the concentration of higher RSRP goes towards lower $A5_{th2}$ and higher values of $A5_{th1}$. This observation provides insight that if larger TTT is used (i.e., dense urban area where mobility is slow), to maintain good RSRP for the users, a higher value of $A5_{th1}$ and a lower value of $A5_{th2}$ should be used. This will ensure that even if TTT is longer; the conditions to fulfill HO will be relaxed, avoiding instances of delayed HO resulting in good RSRP values.

Fig. \ref{fig:GS} shows a 2D plot of mean RSRP versus $A5_{th1}$ and $A5_{th2}$ for TTT of 64ms and 512ms. In this figure, we highlight with a red box the A5 parameter values used as gold standards (GS) by the leading operators in United States. We have also highlighted the blue area where the highest average RSRP has been observed for the analyzed scenario. This comparison shows a significant overlap between the GS and our values of $A5_{th1}$ and $A5_{th2}$ for TTT of 64. However, the location of the blue box changes when TTT is 512ms i.e., optimal values of A5 thresholds change. Therefore, current GS based fixed value setting approach is not optimal and hence the need for self-optimization solution as proposed in this study.


\begin{figure}[t]
\centerline{\fbox{\includegraphics[width=0.48\textwidth,height=2.1in]{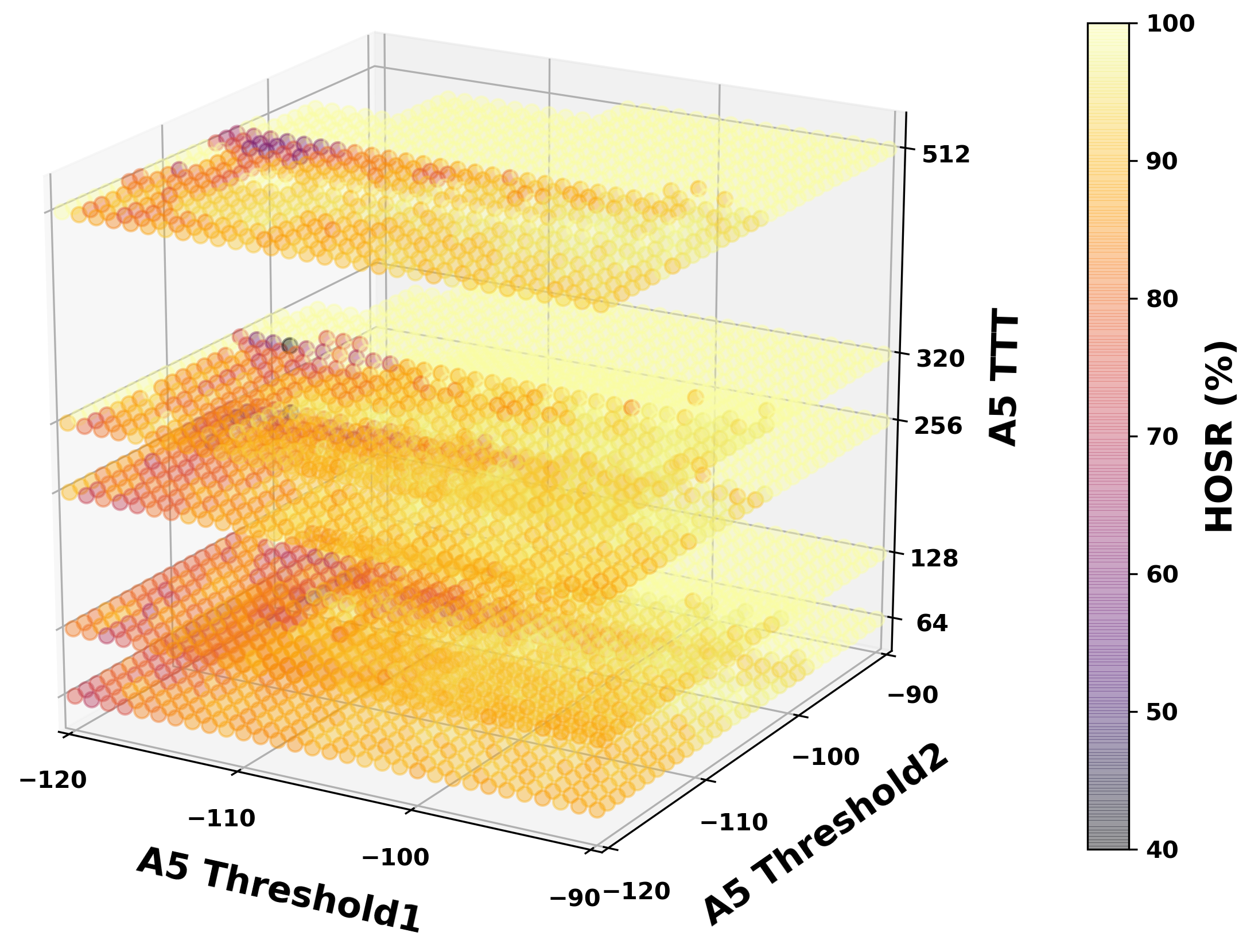}}}
\caption{Impact of A5 thresholds and TTT on HOSR.}
\label{fig:HOSR_Thres1vsThres2vsTTT_new}
\vspace{-0.1in}
\end{figure}




\subsection{Impact on HOSR}
The impact of different A5 thresholds and TTT setting on HOSR is shown in Fig. \ref{fig:HOSR_Thres1vsThres2vsTTT_new}. At first glance these results give the impression that 100\% HOSR can be achieved by using higher values of $A5_{th2}$ (i.e., greater than -100dBm). However, this does not necessarily mean higher $A5_{th2}$ is the optimal setting. As HO conditions using higher $A5_{th2}$ are more challenging to achieve, very few handovers will occur leading to extremely poor RSRP as seen in Fig. \ref{fig:RSRP_Thres1vsThres2vsTTT_new}. In fact, using extreme thresholds and TTT values results in no HO at all. Although these settings result in lower HO failure, the users are forced to stay under inadequate RSRP coverage for a long period leading to poor SINR, throughput, and increased chances of RLF. This can be also validated from Fig. \ref{fig:RSRP_Thres1vsThres2vsTTT_new}, showing the worst mean RSRP in the same area where the HOSR is the highest. Fig. \ref{fig:HOSR_Thres1vsThres2vsTTT_new} also shows that most HO failures occur when lower $A5_{th1}$ is used. This result is expected as poor RSRP of the serving BS is one of the main reasons for HO failure.

The conflicting trend between the results in Fig. \ref{fig:RSRP_Thres1vsThres2vsTTT_new} and Fig. \ref{fig:HOSR_Thres1vsThres2vsTTT_new} shows that there is a trade-off between maximizing RSRP and maximizing HOSR, necessitating the joint optimization of the two KPIs together as proposed in this paper. In the following section, we perform the Sobol index-based sensitivity analysis of the two KPIs with respect to the three cops of interest to gather more insights that can enable joint optimization of RSRP and HOSR.


\section{Machine Learning Models for COP-KPI Relationship}
\label{sec:ML}
\begin{figure}[t]
\centerline{\fbox{\includegraphics[width=0.48\textwidth,height=2.1in]{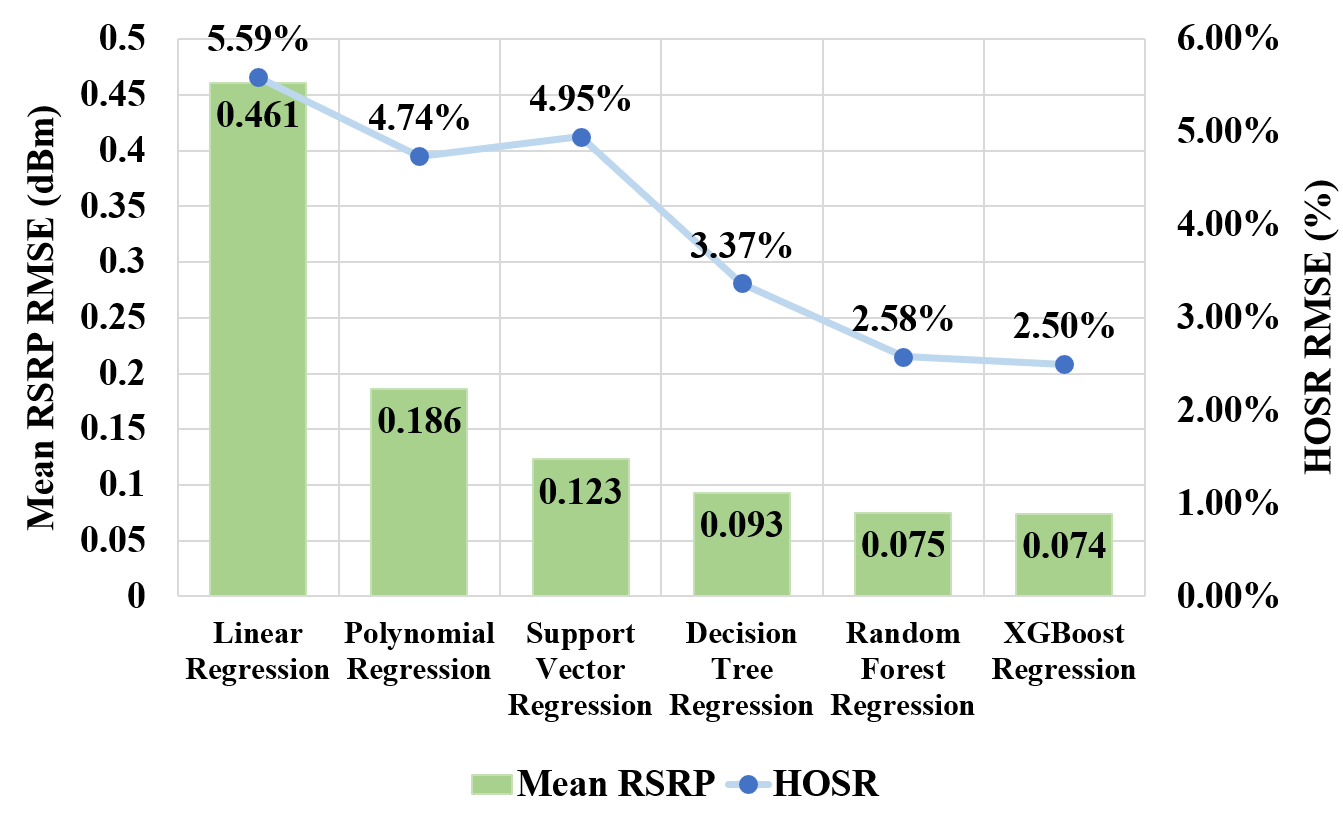}}}
\caption{Comparison of different machine learning algorithms for RSRP and HOSR prediction.}
\label{fig:Results_new}
\end{figure}

This section presents the performance of machine learning algorithms in quantifying the COP-KPI relationship. $A5_{th1}$, $A5_{th2}$, and TTT of A5 are used to predict the network performance in terms of mean RSRP and HOSR. A 80\%-20\% train-test data split is used and the performance of six different regression techniques is evaluated. Fig. \ref{fig:Results_new} shows the performance of each algorithm in terms of root mean square error (RMSE). Due to the complex non-linear relationship between COPs and KPIs, linear regression is not able to capture the relationship leading to a high RMSE of 0.461dBm and 5.59\% for mean RSRP and HOSR prediction, respectively. Similarly, fourth order polynomial and support vector regression techniques also failed to capture the COP-KPI relationship displaying higher RMSE compared to other algorithms. Results also show that tree-based algorithms exhibit promising results in predicting the KPIs. Top 3 algorithms with lowest RMSE for both RSRP and HOSR are all tree-based with XGBoost being the best showing RMSE of only 0.074dBm and 2.5\% for mean RSRP and HOSR, respectively. Overall, results show that machine learning algorithms can capture the relationship between COPs and KPIs with fairly low error. 

\begin{figure}[t]
\centerline{\fbox{\includegraphics[width=0.48\textwidth,height=2.1in]{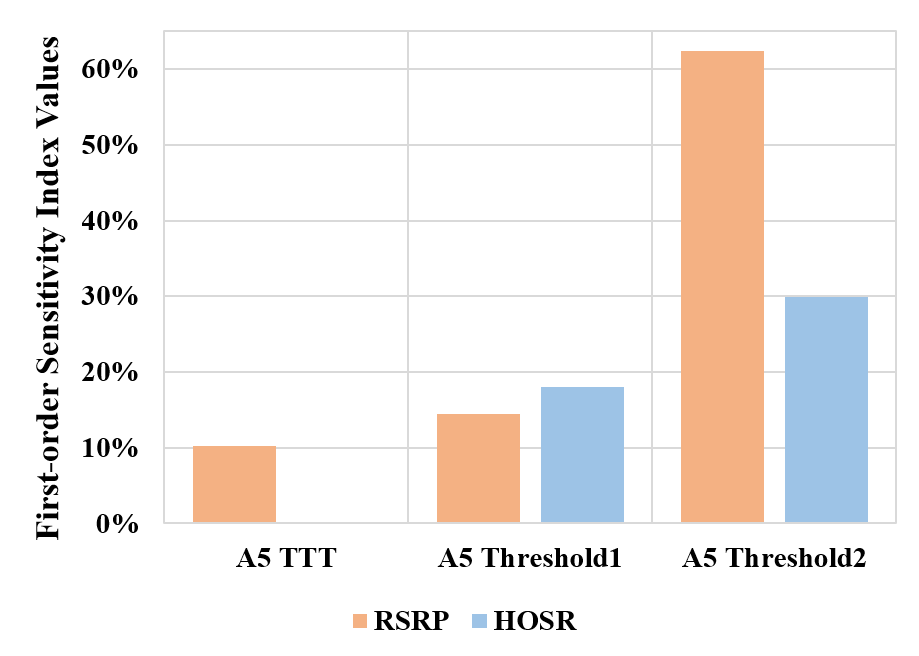}}}
\caption{Sobol sensitivity analysis.}
\label{fig:sobol}
\vspace{-0.1in}
\end{figure}

The relative effect of each of the three COPs on the two KPIs using Sobol based variance sensitivity analysis method~\cite{SOBOL2001271} is plotted in Fig. \ref{fig:sobol}. It is observed that $A5_{th2}$ has the largest impact on the performance of both mean RSRP and HOSR followed by $A5_{th1}$. This shows that a small variation in $A5_{th2}$ will have a large impact on both KPIs. Meanwhile, TTT has almost no effect on HOSR while it has some impact on mean RSRP. This shows that TTT can be varied to optimize RSRP without significant degradation in HOSR. The conflict between both KPIs can be avoided to some extent by varying only TTT to optimize RSRP. In addition to avoiding parametric conflict in existing SON functions \cite{lateef2015lte}, these insights are useful for the operators while tuning the parameters of event A5.

\section{Objective Function Optimization}
\label{sec:optimization}
\begin{figure}[t]
\centerline{\fbox{\includegraphics[width=0.48\textwidth,height=2.1in]{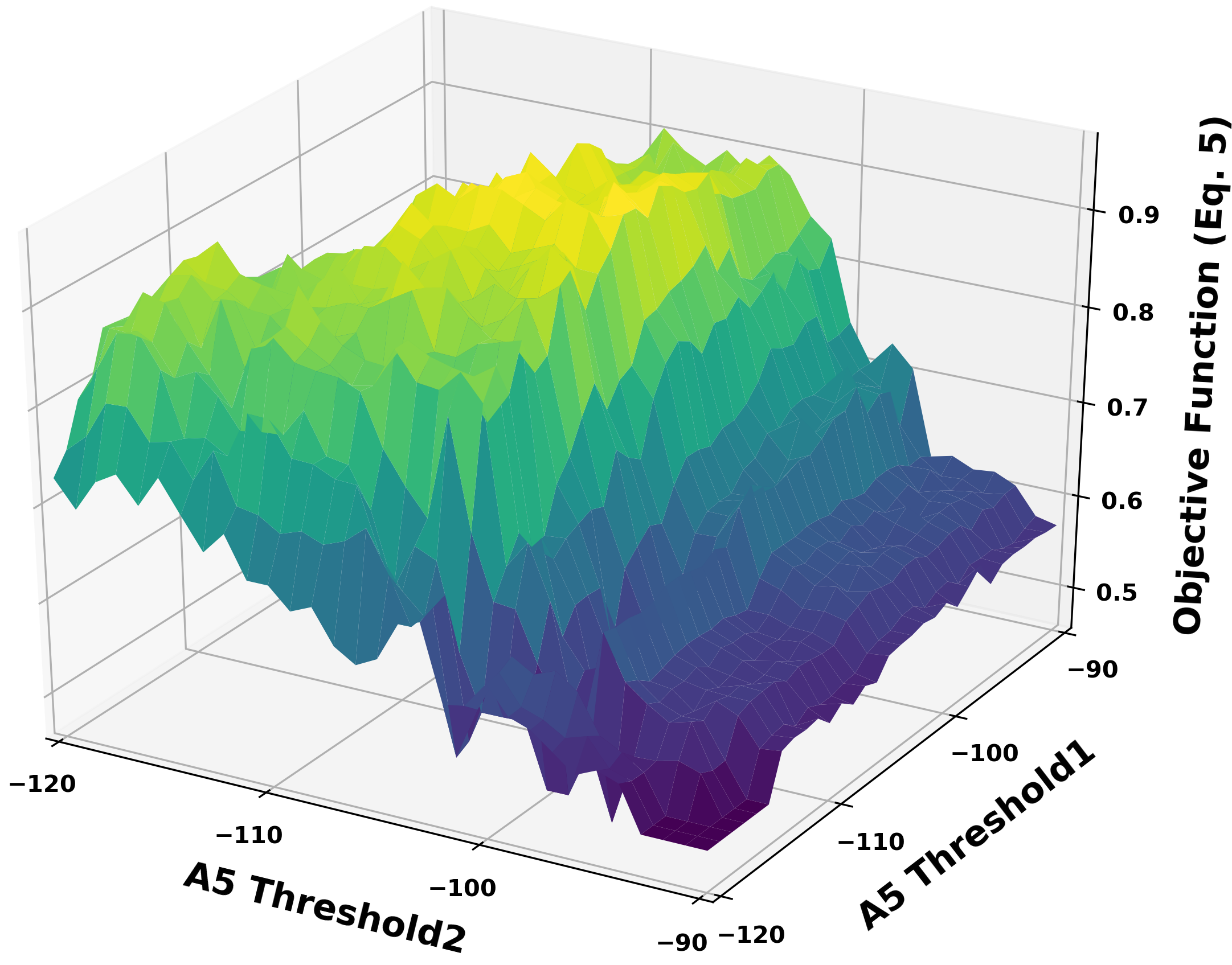}}}
\caption{Objective function defined in eq. \eqref{eq:optimization} with  $\alpha=0.5$ and TTT=64ms.}
\label{fig:Obj_thres1vsthres2}
\end{figure}

Fig. \ref{fig:Obj_thres1vsthres2} shows the plot of the objective function defined in eq. \eqref{eq:optimization} with 0.5 value of $\alpha$. This plot shows how the objective function varies with changes in thresholds and fixed TTT of 64ms. As shown in the plot, there are several maxima located at around -90dBm to -100dBm for $A5_{th1}$ and -110dBm to -120dBm for $A5_{th2}$. It can be seen that \eqref{eq:optimization} is a non-convex optimization problem. This non-convex problem can be solved either through brute force search or heuristic solutions. We compare the performance of brute force method for optimization with well-defined heuristic approach, genetic algorithm (GA). Table \ref{table:GA} shows that the GA can converge 48 times faster compared to brute force method. The fast convergence time especially can make the solution agile for fast changing network conditions. Brute force guarantees optimal value but is not computationally viable particularly for large scale problems i.e. ones involving multiple parameters and cells. In most cases, GA can converge quickly to near optimal values.

\begin{table} [!t]
\centering
\caption{{Comparison between Genetic Algorithm and Brute Force}}
\label{table:GA}
\begin{tabular}{|>{\centering\arraybackslash} m{1.3cm} | >{\centering\arraybackslash}m{1.1cm}|>{\centering\arraybackslash} m{1.6cm} | >{\centering\arraybackslash} m{0.7cm}| >{\centering\arraybackslash} m{1.4cm}|}
\hline
& \textbf{Objective Function} & \textbf{Mean RSRP (dBm)} & \textbf{HOSR (\%)} & \textbf{Number of Iterations} \\ \hline
\textbf{Genetic Algorithm} & 0.9391 & -109.47 & 92.44 & 100\\ \hline
\textbf{Brute Force} & 0.9709 & -109.34 & 94.17 & 4806\\ \hline
\end{tabular}
\vspace{-0.1in}
\end{table}

A comparison of the best values from GA and brute force with $\alpha=0.5$ is shown in Table \ref{table:GA}. It can be seen that the value of objective function returned by GA is slightly sub-optimal compared to that returned by brute force. There is a minor difference of 0.13dBm and 1.73\% in optimal values of mean RSRP and HOSR between GA and brute force. This difference has two reasons, the sub-optimal convergence of GA and the prediction error of the ML algorithm described in Section \ref{sec:ML}. The optimal values of [TTT, $A5_{th1}$, $A5_{th2}$] through brute force are [128ms, -104dBm, -110dBm] compared to [128ms, -103dBm, -109dBm] through GA. The very small difference in the optimal values of KPIs and COPs verifies the presented solution, that combines ML for KPI prediction with heuristic optimization, can enable self-optimization of A5 and other similar parameters. Compared to gold standard such self-optimization can improve the KPIs like RSRP and HOSR substantially.

\section{Conclusion}
\label{sec:conc}
In the wake of densification and multi-band operation envisioned for 5G and 6G, inter-frequency handover can become major bottle neck in user experience. This paper presents the first solution to systematically analyze and optimize three key mobility management COPs that dictate inter-frequency handover: $A5_{th1}$, $A5_{th2}$ and TTT of event A5. The proposed optimization solution jointly maximizes two KPIs: RSRP and HOSR. As tractable analytical modeling is not a viable approach due to the complexity of the system level dynamics, a data driven approach is leveraged to solve the problem. To address the shortage of data, synthetic data from 3GPP compliant simulator that has been validated against real network data is used. Several state-of-the-art machine learning techniques are used to develop and test COP-KPI model. XGboost can predict mean RSRP and HOSR with RMSE of 0.074dBm and 2.5\%, respectively. Sobol sensitivity analysis shows that $A5_{th2}$ has the highest impact on both RSRP and HOSR while TTT has the least impact on RSRP and almost no impact on HOSR. We solve the joint RSRP and HOSR problem using GA. Results show that proposed data driven modeling and GA based solution is 48 times faster compared to brute force search at the cost of 0.13dBm and 1.73\% reduction in RSRP and HOSR, respectively.

\section*{Acknowledgment}
This work is supported by the National Science Foundation under Grant Numbers 1718956 and 1730650 and Qatar National Research Fund (QNRF) under Grant No. NPRP12-S 0311-190302. The statements made herein are solely the responsibility of the authors. For more details about these projects please visit: http://www.ai4networks.com

\ifCLASSOPTIONcaptionsoff
  \newpage
\fi

\bibliographystyle{ieeetr}
\bibliography{globecomFormat}

\end{document}